\documentclass{article}
\usepackage{cite}
\usepackage{arxiv}
\usepackage[english]{babel}
\usepackage[T1]{fontenc}
\usepackage{lmodern}
\usepackage{mathtools}
\usepackage{relsize}
\usepackage{amssymb}
\usepackage{dsfont}
\usepackage{mathptmx}

\begin{document}
\title{On finite representation of dimensionally regularized one-loop integrals}
\author{Juuso Österman\thanks{juuso.s.osterman@helsinki.fi}\\
  University of Helsinki, Helsinki Institute of Physics}
\maketitle
\begin{abstract}
Dimensional regularization of Euclidean momentum space integrals is a highly successful technique in renormalization of quantum field theories. While it yields a straightforward algorithmic method, with which to evaluate diagrams beyond tree level, the actual integrals can be highly divergent, at least in a traditional sense. In particular, standard one-loop integrals can be expressed in terms of an explicit formula, which associates both ultraviolet and infrared divergent parameter values to analytically continued special function expressions. We aim to discuss the formulation of finite integral expressions corresponding to the analytically continued structures. Effectively, we wish to establish conditions which form an equivalence class for this analytical continuation, or rather form a proper set/conditions of regularization techniques leading to it. This is further demonstrated by considering both partially and fully successful strategies side-by-side, with major emphasis on the two simplest functioning schemes: Gaussian and cut-off regularization. By explicit computations we aim to associate these generalisations of the initial integrals with the results from dimensional regularization, considering both multiple mass (or momentum) scales as well as scaleless cases. We achieve the finiteness of the sought-after integrals by applying one of these suitable schemes along with an additional scheme related scale. This enables us to devise a proper representation of the dimensionally regularized expressions (or a local description) through an operator removing all excess terms with the additional scale(s).    
\end{abstract}
\section{Introduction to convergence in one-loop diagrams}
\noindent
In this article we seek the essence of dimensional regularization \cite{Veltman, Bollini} through the simplest loop level integrals, while neglecting the mathematical aspects arising from the full renomalization process (discussed e.g. in \cite{cicuta1, cicuta2, Breitlohner, Smirchet1, Smirchet2}). While gauge fields theories add complexity through tensor algebra, this is independent of actual (radial) regularization process \cite{Veltman, hooft2, collins, peskin, schwarz}. Thus, we find all relevant information by generalizing the power structures of the integrands found in the loop level corrections of scalar field theories. The computations to follow revolve around absolutely convergent descriptions of relevant integrals (suitably regulated). Using somewhat similar mindset, dimensionally regularized Feynman amplitudes have been discussed in $\alpha$ representation in \cite{Smirchet3}.

In further sections, we introduce specific regulators to loop integrals in order to be able discuss absolutely convergent integrals. However, let us first examine the simplest relevant integrals structures without any such generalizations, and recognize the challenges arising from attempts of exact evaluation of the integrals (with specific parameter values). The simplest $d$-dimensional massive vacuum loop diagram is given by
\begin{equation}
\label{eq:d-1loop1}
\begin{split}
I(m) &= \int \frac{d^d p}{(2 \pi)^d} \frac{1}{p^2+m^2}\\
&\equiv \int_p \frac{1}{p^2+m^2},
\end{split} 
\end{equation} 
where we assume the quadratic mass scale to be strictly positive ($m^2 > 0$) and typically set the space to be three- or four-dimensional. In either traditional dimension, the integral given above is divergent in the ultraviolet (UV) region, as can be seen by considering the radial integrand
\begin{equation}
\frac{p^{d-1}}{p^2+m^2} \underset{p \rightarrow \infty}{\longrightarrow} p^{d-3}\left[ 1 + \mathcal{O} \left( \frac{m^2}{p^2}\right) \right].
\end{equation} 
While this integral is obviously convergent in the infrared (IR) region even in the vanishing mass scale limit, by differentiating it with respect to square of the mass, we find an integrand such that 
\begin{equation}
\frac{p^{d-1}}{(p^2+m^2)^2} \underset{m^2 \rightarrow 0}{\longrightarrow} p^{d-5}\left[ 1 + \mathcal{O} \left( \frac{m^2}{p^2}\right) \right].
\end{equation}
The related integral rather obviously diverges at low values of loop momentum, either inversely or logarithmically. This is of course treated by considering solely non-vanishing mass scales. Thus, defining a master integral of sorts
\begin{equation}
I_\alpha^d (m) = \int \frac{d^d p}{(2 \pi)^d} \frac{1}{(p^2+m^2)^\alpha},  
\end{equation}
we can split the parameter space in two halves (for non-vanishing masses), the UV convergent and divergent cases. The former can be found in straightforward manner by only considering the asymptotical expansions of the integrand that are convergent, i.e. $2 \alpha-d > 0$. This leads to the following straightforward result through application of an integral representation of Euler beta function
\begin{equation}
\label{eq:d-1loop}
\begin{split}
I_\alpha^d (m) & =  \frac{(m^2)^{\frac{d}{2}-\alpha}}{(2\pi)^d} \Omega_d \int_0^\infty dy \frac{y^{d-1}}{(y^2+1)^\alpha}\\
&= \frac{2(m^2)^{\frac{d}{2}-\alpha}}{2(4 \pi)^\frac{d}{2} \Gamma \left(\frac{d}{2}\right)} \int_0^\infty dz \frac{z^{\frac{d}{2}-1}}{(z+1)^\alpha}\\
&=  \frac{(m^2)^{\frac{d}{2}-\alpha}}{(4 \pi)^\frac{d}{2} }
\frac{\Gamma \left( \alpha-\frac{d}{2} \right)}{ \Gamma (\alpha)},\\
\end{split}
\end{equation}
where we denote the area of $d$ dimensional unit sphere as $\Omega_d = \frac{2 \pi^\frac{d}{2}}{\Gamma \left(\frac{d}{2} \right)}$ \cite{peskin, schwarz}. For integer valued dimensions, this result follows from a trivial Gaussian integral such that
\begin{equation}
\label{eq:axiomgauss}
\begin{split}
\int d^d x e^{-x^2} &= \frac{\Omega_d}{2} \int_0^\infty dy y^{\frac{d}{2}-1} e^{-y} =  \frac{\Omega_d}{2} \Gamma \left(\frac{d}{2} \right)\\
&= \left[\int_{-\infty}^\infty dz e^{-z^2} \right]^d = \pi^\frac{d}{2}.
\end{split}
\end{equation}
By axiomatically extending the Gaussian integral result to non-integer (positive) dimensions, we find the formula given in equation (\ref{eq:d-1loop}) to apply exactly to all parameters with both $d >0$ and $d-2\alpha < 0$. 
\section{Dimensional regularization of standard Euclidean integrals}
In general we choose to keep the dimension of the space-time arbitrary, other than take it to be explicitly positive. Typically it is set to either three or four \cite{Veltman, schwarz, peskin}. This obviously leads to the traditional region of convergence, $\alpha > \frac{d}{2}$, to not contain the very first integral given in equation (\ref{eq:d-1loop1}). However, the expression on the right-hand side of equation (\ref{eq:d-1loop}) can be extended close to these parameter values, through standard analytical continuation of Euler gamma functions. Still, convergence is limited to $\frac{d}{2}-\alpha \not \in \mathbb{Z}_{-} \cup 0$, which is most easily resolved by the name sake of the regularization scheme. Specifically, to ascertain the convergence, we exclusively (tend to) consider non-integer valued dimensions instead, setting $d = n - 2\epsilon$, with $n \in \mathbb{N}$. These two steps form the core of dimensional regularization in standard computations \cite{collins, peskin, schwarz}.  

However, this technique does not limit to such parameter values, rather, we can interpret the result for any case in which the special function in the numerator does not diverge. Thus, instead of small deviations from positive integers, we are able to consider parameter space $\left\{\alpha-\frac{d}{2}, d \right\} \in \left[\mathbb{R} \setminus \left(\mathbb{Z}_- \cup 0 \right)\right] \times \left( \mathbb{R}_+ \setminus \mathbb{N} \right)$. Thus, the result limits out only a countable set of parameters (rather than previously uncountable due to the infinities). As a minor point of interest, we note that the standard one-loop formula, as given in dimensional regularization, actually allows the use of negative dimensions. However, as the Gaussian integral, which acts as the basis of our expression, is only convergent for strictly positive dimensions, we choose to remain in the positive real axis. 
\subsection{Characteristics}
Specifically, dimensional regularization takes the finite solution of the convergent region as given in equation (\ref{eq:d-1loop}) and analytically continues the gamma functions to negative real axis. Thus, we have an expression that describes almost all of parameter space. Hence, in terms of dimensional regularization, the integrals with non-converging parameters exist only through injective mapping to corresponding gamma functions. As such, a proper interpretation in terms of these standard one loop integrals is more of a correspondence between two indices, dimension $d$ and power $\alpha$, the Euclidean quadratic mass parameter, $m^2$, and the final special function result:
\begin{equation}
\left\{m^2, d, \alpha \right\} \longmapsto \frac{(m^2)^{\frac{d}{2}-\alpha}}{(4 \pi)^\frac{d}{2} }
\frac{\Gamma \left( \alpha-\frac{d}{2} \right)}{ \Gamma (\alpha)}. 
\end{equation} 

This structure of solutions carries over certain algebraic properties of the original integral representation, inherently corresponding to supposed sufficient convergence. In particular we are allowed to commute partial derivatives and integral operators at will, enabling versatile manipulation of the integrand \cite{kleinert, Smirchet3}. By direct computation on both the integrand, as well as the results it maps to, we can confirm that differentiation with respect to mass scale effectively lowers the index corresponding to the propagator by one $\alpha \mapsto \alpha+1$:
\begin{equation}
-\partial_{m^2} \int_p \frac{1}{(p^2+m^2)^a} = a  \int_p \frac{1}{(p^2+m^2)^{a+1}}.
\end{equation} 
This already implies that the nature of the mapping is to isolate well-behaving blocks of the original integral. 

While the mass scale is implicitly required to be non-vanishing, through suitable commutation relations and assumed convergence, we can find insight on  on the scaleless behaviour dimensionally regularized one-loop integrals. Let us consider the following integral for arbitrary dimension $d$ (e.g. $d = 4 - 2 \epsilon$), with the help of Feynman parametrization:
\begin{equation}
\begin{split}
\int_p \frac{1}{p^2 (p^2+m^2)} &= \int_0^1 dx \int \frac{d^d p}{(2\pi)^d} \frac{1}{[p^2+ xm^2]^2}\\
&= \frac{\Gamma \left(2-\frac{d}{2} \right) (m^2)^{\frac{d}{2}-2}}{(4\pi)^\frac{d}{2} \Gamma (2)} \int_0^1 dx x^{\frac{d}{2}-2}\\
&= -\frac{1}{m^2} \frac{\Gamma \left( 1- \frac{d}{2} \right) (m^2)^{\frac{d}{2}-1}}{(4\pi)^\frac{d}{2}}\\
&= - \frac{1}{m^2}\int_p \frac{1}{p^2+m^2},
\end{split}
\end{equation}
where we changed the order of integration, effectively supposing that the result converges \cite{peskin, kleinert}. Of course this does not hold for any $d < 2$, in traditional manner, but we apply the same reasoning as used for the initial one-loop result. By following up with straightforward linear algebra on the integral, we find 
\begin{equation}
\int_p \frac{1}{p^2 (p^2+m^2)} = \frac{1}{m^2} \left[\int_p \frac{1}{p^2}- \int_p\frac{1}{p^2+m^2} \right].
\end{equation}
Thus, we can state that in dimensional regularization, we can consider all integrals over generalized monomials to vanish (for any arbitrary dimension $d$ and parameter $\alpha$) such that 
\begin{equation}
\int  \frac{d^d p}{p^\alpha} = 0,
\end{equation}
which is known as Veltman's identity in literature \cite{collins, kleinert}, or more broadly Hadamard regularization in mathematics \cite{hadamard}. Veltman's identity tends to be stated usually through homogenity or by using dimensional scaling arguments (e.g. \cite{smirnov}):

\begin{equation}
\int  \frac{d^d p}{p^\alpha} = \int \frac{d^d (xp)}{(xp)^\alpha} = x^{d-\alpha} \int \frac{d^d p}{p^\alpha},
\end{equation}
where the additional $x$ represents an arbitrary scale.

In essence, dimensional regularization is injective interpretation between divergent integrals and gamma function representation of the convergent parameter region. Due to the supposed convergent nature, this structure is taken to allow commutation with both derivative and integral operators (the latter in particular allowing the application of re-parametrizations) \cite{Smirchet3}.

 As a next step we seek to generalize the original integrals to allow explicit computation in all of the sought-after parameter space given earlier. In addition, we want the structure to allow straight-forward isolation of dimensionally regularized results, further still allowing us to observe qualities listed above. Our aim is to establish a consistent convention, or set of rules, with which to deal with generic parameter values in non-trivial multi-loop/propagator computations. In these the convergent region of parameter space can be seen to with the help of well-established special functions, while the divergent region may lead to difficult-to-see 'analytical continuations'.

\section{Regularizations with incomplete parameter space}
Although the one-loop integral structure is somewhat simple, not every regularization yields us the results we are after. In this section, we want to explicitly demonstrate through two examples, how strict demands the convergence conditions are.

Let us limit ourselves to the IR finite master integrals. Thus, the convergence, or divergence, is established through the beta function type integral such that 
\begin{equation}
\mathcal{B} \left(\frac{d}{2}, \alpha-\frac{d}{2} \right) = \int_0^\infty dt \frac{t^{\frac{d}{2}-1}}{(1+t)^\alpha}.
\end{equation} 
A natural way to approach the challenges is to adjust dimension with a Mellin type monomial regulator $t^z$ \cite{hawking}, with an arbitrary parameter $z$ (or rather arbitrarily small), as to write 
\begin{equation}
\int_0^\infty dt \frac{t^{\frac{d}{2}+z-1}}{(1+t)^\alpha} = \mathcal{B} \left(\frac{d}{2}+z, \alpha-z-\frac{d}{2} \right).
\end{equation}  
This approach does indeed enable finite evaluation of each integral with $\alpha > 0$. However, for any
$\alpha = -\beta  \leq 0$ we find the integral
\begin{equation}
\int_0^\infty dt t^{\frac{d}{2}+z-1} (1+t)^\beta \rightarrow \infty
\end{equation}
to explicitly diverge for all $z$ either in the small loop momentum region, or large momentum region. Thus, Mellin regularization effectively works only for half of parameter space. However, it is noteworthy to mentally associate this regularization strategy to minor shift from a divergent dimension to convergent dimensional region, which is exactly how the most common results in dimensional regularization are expanded \cite{peskin, schwarz, hooft2}.

Another interesting approach is to insert a higher order polynomial in place of the propagator, again with an arbitrary extra scale. For a simple three-dimensional one-loop integral this would correspond to
\begin{equation}
\int_0^\infty dx \frac{ x^2}{x^2+1} \longmapsto  \int_0^\infty dx \frac{x^2}{a x^4 +x^2+1}
\end{equation} 
for some positive (yet small) $a$. This specific case can decomposed quite a bit further by denoting $w_{\pm} = \frac{1\pm \sqrt{1-4 a}}{2a}$. Now we find
\begin{equation}
\begin{split}
\int_0^\infty dx \frac{x^2}{a x^4 +x^2+1} &= \frac{1}{a(w_+ - w_-)} \int_0^\infty dx \left[\frac{w_+}{x^2+w_+}-\frac{w_-}{x^2+w_-} \right]\\
&= \frac{\pi}{2a} \frac{\sqrt{w_+}-\sqrt{w_-}}{w_+-w_-}\\
&= \frac{\pi}{2\sqrt{2a}} \frac{\sqrt{1+\sqrt{1-4a}}-\sqrt{1-\sqrt{1-4a}}}{\sqrt{1-4a}}\\
&= \frac{\pi}{2 \sqrt{a}} \left( 1- \sqrt{a}+\frac{3a}{2}-\frac{5 a^\frac{3}{2}}{2}... \right).
\end{split}
\end{equation}
Thus we would indeed be able to isolate the sought-after beta function structure by removing the $a$ dependence from the equation as 
\begin{equation}
\begin{split}
\int_0^\infty dx \frac{x^2}{x^2+1} &= \frac{1}{2} \int_0^\infty dt \frac{t^{\frac{1}{2}}}{1+t}\\
&\longmapsto \frac{\Gamma \left( \frac{3}{2} \right) \Gamma \left(-\frac{1}{2} \right)}{2}\\
&= -\frac{\pi}{2}.
\end{split}
\end{equation}
However, this methodology fails in producing a convergent structure for any case with non-positive $\alpha$ in the radial integral, similar to the Mellin approach. While these strategies proved unfruitful, they yield some insight to the relevant strategies. First, convergence is achieved through the introduction of a new scale. Second, UV convergence fails if the introduced regularization does not suppress the asymptotic growth faster than polynomials, or if the introduced method introduces a new source of divergence.

\section{Finite UV regularizations}
\noindent
In this section we discuss the two simplest means to sufficiently suppress polynomial growth: step function cut-off \cite{jaffe, peskin, schwarz} and exponential decay. These two types of regulators succeed where the two presented above failed, in addition to allowing straightforward extraction of the results found through dimensional regularization. These initial computations takes place with Euclidean (positive) mass scales (as well as the scales corresponding to either scheme).

\subsection{Cut-off regularization}
Momentum shell cut-off is even more traditional approach than dimensional regularization \cite{peskin, jaffe, schwarz}. While some correspondence between different relations have been mapped, we aim to keep computations more general than is traditional and through that find a meaningful correspondence between the two schemes. Technically, we introduce a radial cut-off, $\theta \left(K - p\right)$, through the standard Heaviside step function and some large value $K$. As the integrand is chosen to be IR finite, and without poles, we find the integral to be clearly bound above such that 
\begin{equation}
\begin{split}
\int_p \frac{\theta \left(K-p \right)}{(p^2+m^2)^\alpha} \leq  \frac{2 K^d}{(4 \pi)^\frac{d}{2} d \Gamma \left( \frac{d}{2} \right)  } \underset{0\leq p \leq K}{\text{max}} \left[\frac{1}{(p^2+m^2)^\alpha} \right]. 
\end{split}
\end{equation}
Thus, we can safely consider each and any parameter value we wish. The adjusted integral can also be calculated explicitly in terms of special functions. The hypergeometric representation of the regularized integral reads
\begin{equation}
\begin{split}
\tilde{I}_\alpha^d (K,m) &= \int_p \frac{\theta \left(K-p \right)}{(p^2+m^2)^\alpha}= \frac{1}{(2\pi)^d} \int d\Omega_d \int_0^{K} dp \frac{p^{d-1}}{(p^2+m^2)^\alpha}\\
&= \frac{ K^d}{(4 \pi)^\frac{d}{2}\Gamma \left( \frac{d}{2} \right)} \int_0^1 \frac{dw w^{\frac{d}{2}-1}}{(K^2 w + m^2)^\alpha}\\
&=\frac{ K^d}{(4 \pi)^\frac{d}{2}\Gamma \left(1+ \frac{d}{2} \right)(m^2)^\alpha}   {}_2 F_1 \left[\alpha,\frac{d}{2}, 1+ \frac{d}{2}, -\frac{K^2}{m^2} \right].
\end{split}
\end{equation}
However, as the common power series expansion of hypergeometric function takes place exclusively around small values of variables (as opposed to large negative values), it is beneficial to explicitly perform the expansion suitably on the integrand. This allows us to better associate non-standard power series expansions to the relevant special functions. By returning to the second to last row of the previous expression, we can write
\begin{equation}
\begin{split}
K^d\int_0^{1}  dw \frac{w^{\frac{d}{2}-1}}{\left(K^2 w + m^2 \right)^\alpha} &= K^{d-2\alpha} \int_{\frac{m^2}{K^2}}^{1} dw \frac{w^{\frac{d}{2}-\alpha-1}}{\left(1 + \frac{m^2}{K^2 w} \right)^\alpha} + m^{d-2\alpha} \int_0^1 dv \frac{v^{\frac{d}{2}-1}}{\left(1+v \right)^\alpha}\\
&= K^{d-2\alpha} \sum_{n= 0}^\infty {-\alpha \choose n} \left( \frac{m^2}{K^2}\right)^n \int_{\frac{m^2}{K^2}}^1 dw w^{\frac{d}{2}-\alpha-n-1}\\
&+ m^{d-2\alpha} \sum_{n= 0}^\infty {-\alpha \choose n} \int_{0}^1 dv v^{\frac{d}{2}+n-1}\\
&= K^{d-2\alpha} \sum_{n= 0}^\infty {-\alpha \choose n} \left( \frac{m^2}{K^2}\right)^n \frac{1}{\frac{d}{2}-\alpha-n} \\
&+ m^{d-2 \alpha} \sum_{n=0}^\infty {-\alpha \choose n} \left( \frac{1}{\frac{d}{2}+n}-\frac{1}{\frac{d}{2}-\alpha-n} \right).
\end{split}
\end{equation}
Here we note that a convergent Euler beta function ($x, y >0$) is given by the following series
\begin{equation}
\begin{split}
\mathcal{B}(x,y) &= \int_0^\infty dt \frac{t^{x-1}}{(1+t)^{x+y}}\\
&= \int_0^1 dt \frac{t^{x-1}}{(1+t)^{x+y}} + \int_0^1 dv \frac{v^{y-1}}{(1+v)^{x+y}}\\
&= \sum_{n=0}^\infty {-(x+y) \choose n } \int_0^1 dt \left( t^{x+n-1}+t^{y+n-1} \right)\\
&= \sum_{n=0}^\infty {-(x+y) \choose n } \left( \frac{1}{x+n}+\frac{1}{y+n} \right).
\end{split}
\end{equation}
This series representation of beta function converges for parameter values beyond postive real axis, such that $x, y \not \in \mathbb{Z}_- \cup 0$. Hence, we can recognize the full expansion for the large UV cut-off as 
\begin{equation}
\label{eq:d-1loopuv}
\begin{split}
\tilde{I}_\alpha^d (K, m) &= \frac{K^{d-2\alpha}}{(4 \pi)^\frac{d}{2} \Gamma \left( \frac{d}{2} \right)} \sum_{n= 0}^\infty {-\alpha \choose n} \left( \frac{m^2}{K^2}\right)^n \frac{1}{\frac{d}{2}-\alpha-n}  + \frac{(m^2)^{\frac{d}{2}-\alpha}}{(4 \pi)^\frac{d}{2} }
\frac{\Gamma \left( \alpha-\frac{d}{2} \right)}{ \Gamma (\alpha)}\\
&= \frac{K^{d-2\alpha}}{(4 \pi)^\frac{d}{2} \Gamma \left(\frac{d}{2} \right)} \left[\frac{1}{\frac{d}{2}-\alpha} + \mathcal{O} \left( \frac{m^2}{K^2} \right)\right] + I_\alpha^d (m),
\end{split}
\end{equation}
where the gamma functions have been suitably analytically continued to accept negative parameter values. This result explicitly relates to the one-loop dimensionally regularized solution, with the additional scale (source of divergence) being separated and its leading term being independent of the relevant mass scale. Using this expression, our integral of interest  extends to cover the full parameter space (of interest) apart from a countable set of divergent points. Also, the original convergent result is plain to see by taking the limit $\frac{K}{m} \rightarrow \infty$, for $\frac{d}{2}-\alpha < 0$ and constant mass scale, we find the familiar result 
\begin{equation}
\tilde{I}_\alpha^d (K,m) \longrightarrow I_\alpha^d (m) + \mathcal{O} \left( K^{d-2\alpha} \right).
\end{equation}

It is also noteworthy that this expression specifically introduces the need of the namesake for dimensional regularization, as we can choose specific parameter values not in agreement with this expansion, while the integral is still convergent. A prime example on this is the well known radial integral
\begin{equation}
\begin{split}
\int_0^K \frac{dp p^3}{(p^2+m^2)^2} &= \frac{K^4}{2 m^4} {}_2 F_1 \left[2, 2, 3, -\frac{K^2}{m^2} \right]\\
&=\text{ln} \left(1+ \frac{K^2}{m^2} \right) -\frac{K^2}{m^2+ K^2}\\
&\underset{K \rightarrow \infty}{\longrightarrow} \ln \left(\frac{K^2}{m^2} \right)-1,
\end{split}
\end{equation}
which we can only recognize by setting $d = 4-2 \epsilon$ and noting that 
\begin{equation}
\ln \left( \frac{K^2}{m^2} \right) = \frac{1}{\epsilon} \left(K^{2\epsilon} -m^{2\epsilon} \right).
\end{equation}

While cut-off scheme clearly yields everything we are after, it is in no way unique. Thus, we explore a second straight-forward example, which employs exponential suppression.
\subsection{Gaussian regularization}
As proposed above let us consider the standard integral with Gaussian suppression (modified by some small scale $\delta$). We can again clearly bind this expression from above such that
\begin{equation}
\begin{split}
\int_p \frac{e^{-\delta p^2}}{(p^2+m^2)^\alpha} &\leq \int_p e^{-\delta p^2} \left[\frac{\theta(\alpha)}{m^{2\alpha}} + \frac{\theta (-\alpha)}{2^\alpha m^{2 \alpha}} + \frac{\theta(-\alpha)}{2^\alpha p^{2\alpha}} \right]\\
&= \frac{1}{(4 \pi)^\frac{d}{2}} \left[ \frac{\theta (\alpha)}{ \delta^{\frac{d}{2}} m^{2 \alpha}} + \frac{\theta(-\alpha)}{\delta^\frac{d}{2} 2^{\alpha} m^{2\alpha}} + \frac{\theta(-\alpha)}{2^{\alpha} \delta^{\frac{d}{2}-\alpha}} \frac{\Gamma \left( \frac{d}{2}-\alpha\right)}{\Gamma \left( \frac{d}{2} \right)} \right].
\end{split}
\end{equation}
With this established, we can continue to explicit evaluation of the integral through special functions and corresponding power series expansion. In straightforward manner, we find an expression in terms of confluent hypergeometric function of the second kind (as a function of the additional scale $\delta$) \cite{stegun}
\begin{equation}
\begin{split}
\mathfrak{I}_\alpha^d (\delta, m) &= \int \frac{d^d p}{(2 \pi)^d} \frac{e^{-\delta p^2}}{(p^2+m^2)^\alpha}\\
&=\frac{(m^2)^{\frac{d}{2}-\alpha}}{(4 \pi)^\frac{d}{2} \Gamma \left(\frac{d}{2} \right) } \int_0^\infty dw \frac{ e^{-\delta m^2 w} w^{\frac{d}{2}-1}}{(w+1)^\alpha}\\
&=\frac{(m^2)^{\frac{d}{2}-\alpha} }{(4 \pi)^\frac{d}{2} } U \left[\frac{d}{2}, \frac{d}{2}+1-\alpha, m^2 \delta \right].
\end{split}
\end{equation}
This expression can be related back to confluent hypergeometric function of the first kind such that 
\begin{equation}
U(\alpha, \beta, z) = \frac{\Gamma \left(1-\beta \right)}{\Gamma \left(\alpha+1-\beta \right)} {}_1 F_1 \left[\alpha, \beta, z \right] + \frac{\Gamma \left(\beta-1 \right) z^{1-\beta}}{\Gamma \left(\alpha \right)} {}_1 F_1 \left[\alpha+1-\beta, 2-\beta, z \right],   
\end{equation}
where the small variable $z$ enables the following power series 
\begin{equation}
\begin{split}
{}_1 F_1 \left[\alpha, \beta, z \right] = \sum_{n=0}^\infty \frac{\Gamma \left( \alpha+n \right) \Gamma \left(\beta \right)}{ n! \Gamma (\alpha) \Gamma \left(\beta +n \right)} z^n.
\end{split}
\end{equation}
Thus, the Gaussian result can be written in terms of the following $\delta$ expansion for any $d > 0$ such that
\begin{equation}
\label{eq:d-1loopgaus}
\begin{split}
\mathfrak{I}_\alpha^d (m) &= \frac{(m^2)^{\frac{d}{2}-\alpha}  }{(4 \pi)^\frac{d}{2} }  \left\{ \frac{\Gamma \left(\alpha-\frac{d}{2} \right)}{\Gamma \left(\alpha \right)} \left[1 +\frac{ d m^2 \delta }{d+2-2\alpha} + \mathcal{O} \left(\delta^2 \right) \right] \right. \\
&+\left.   \delta^{\alpha-\frac{d}{2}} \frac{\Gamma \left(\frac{d}{2} -\alpha \right)}{\Gamma \left( \frac{d}{2} \right)} \left[1 +\frac{\alpha m^2 \delta}{\alpha+1-\frac{d}{2}} + \mathcal{O} \left(\delta^2 \right) \right] \right\},
\end{split}
\end{equation}
where the result is given in terms of analytically continued (to negative real axis) Euler gamma functions. Also, we can easily discern the dimensionally regularized result as part of the full expansion, confirmed to function properly by being reached at the limit $\delta \rightarrow 0$ while $\alpha-\frac{d}{2} > 0$. 

Both of the regularization schemes yield non-trivial power series expansion, interpreted through well-known special functions. While they behave well in all of the parameter space, they still yield the sought-after standard results upon suitable limits. 

\section{UV finite mapping}
By looking back to the structure that can be clearly seen in equations (\ref{eq:d-1loopuv}) and (\ref{eq:d-1loopgaus}), we recognize that on the convergent expansions $I_\alpha^d(m)$ corresponds to elements with only mass scale dependence (i.e. no Gaussian regulator nor UV cut-off). The naive idea is to extract mass dependence of said convergent integrals, but it can easily be noted that mass dependence is heavily mixed with other scales in the expansion. Instead we recognize that the function is of the form $C(a,b) = A(a)+B(a,b)$, where all terms in $B(a,b)$ contain explicit $b$ dependence. Now we can construct a proper operator by combining partial derivative and anti-derivative (integration without constant term) to isolate terms $A(a)$. The relevant relation is given by
\begin{equation}
\begin{split}
A(a) &= C(a,b) - B(a,b)\\
&= \left[\mathds{1}-\int db \partial_b \right] C(a,b).
\end{split}
\end{equation}
It is trivial to extend this expression to our dimensionally regularized one-loop integrals. Thus, in terms of finite range integrals we can express the integral of interest as 
\begin{equation}
\left[\mathds{1}-\int dK \partial_{K} \right] \tilde{I}_\alpha^d (K,m) \longmapsto I_\alpha^d(m)
\end{equation}
or alternatively we can relate it to the Gaussian integrals such that
\begin{equation}
\left[\mathds{1}-\int d\delta \partial_{\delta} \right] \mathfrak{J}_\alpha^d (m). \longmapsto I_\alpha^d(m).
\end{equation}
Unlike the traditional interpretation of dimensional regularization, with analytical continuation or divergent monomials, which are neglected, we have written an explicit mapping between two finite expressions. In addition, the mapping takes place through a simple linear operator. This formulation describes explicitly the class of suitable generalizations of the original integral, two members of which have been explicitly listed in previous sections. Hence we have found a proper two-step mapping from the initial (possibly divergent) integral to the master formula. 

While still focusing in the IR finite structure, we have two structurally relevant qualities to consider: integrals with multiple scales and partial derivation with respect to quadratic mass scale(s). First, we want to ascertain that the partial derivative commutes with the introduced operator structure, producing in either case the familiar correspondence between index $\alpha$ and differentiation. It is obvious that either of the UV finite generalizations do not follow the desired relation:
\begin{equation}
\begin{split}
\alpha \tilde{I}_{\alpha+1}^d(K,m) &= \partial_{m^2} I_{\alpha}^d(m) + \mathcal{O} \left(K^{d-2 \alpha} \right)\\
&\neq \partial_{m^2} \tilde{I}_\alpha^d (K,m),
\end{split}
\end{equation}
as obviously the highest order contribution in terms of $K$ must vanish upon differentiation. However, all the excess terms in either expression contain explicit $K$ dependence. Thus, we can write
\begin{equation}
\alpha \left[\mathds{1} - \int dK \partial_K \right] \tilde{I}_{\alpha+1}^d(K,m) = \partial_{m^2}\left[\mathds{1} - \int dK \partial_K \right]  \tilde{I}_{\alpha}^d(K,m),
\end{equation} 
which states the sought-after symmetry.
In the next section,  we introduce (inspired via the derivation of Veltman's identity) a slight generalization with two massive propagators, in effect demonstrating the methods in dimensional regularization and these finite representations side-by-side. Afterwards, we further generalize this expression by including an external finite momentum scale. 
\section{Multiple non-vanishing mass scales}
The computations in this section are performed to all three regularization schemes of interest. We express results in terms of known special function relations when possible, in addition to power series representations. The special function structures for dimensional regularization have been explicitly derived for more versatile kinematic structures (using Mellin-Barnes rerpresentation) in \cite{boos, davydchev3, tarasov, Fleischer, Bluemlein}.

 Another interesting division is between the parts of parameter space that can be honestly re-parametrized using Feynman's technique and the rest requiring expansions using Newton's generalized binomial theorem. The former leads to intuitive and clear calculations, while the latter is at best lengthy in comparison. 
\subsection{Dimensional regularization}
 Let us first define the suitable integral of interest such that the two propagators share the same loop momentum without any external contributions:
\begin{equation}
\label{eq:gen1}
I_{\alpha \beta}^d (m, M) = \int_p \frac{1}{(p^2+m^2)^\alpha (p^2+M^2)^\beta},
\end{equation}
where we assume the following mass hierarchy $M > m > 0$. Supposing first that both parameters are strictly positive, $\alpha, \beta >0$. Thus, the most efficient strategy is to, again, apply Feynman parametrization, and re-write the integrand as
\begin{equation}
\begin{split}
\frac{1}{(p^2+m^2)^\alpha (p^2+M^2)^\beta} &= \frac{\Gamma (\alpha + \beta)}{\Gamma (\alpha) \Gamma (\beta)} \int_0^1 \frac{dx x^{\alpha-1} (1-x)^{\beta-1}}{\left[p^2+x m^2+(1-x)M^2 \right]^{\alpha+\beta}}.
\end{split}
\end{equation}
The simplest way to find a special function solution is to change the order integration and use the known one-loop results on the new structure arising from Feynman parametrization (effectively a change of basis). This is taken to be a well defined operation, although rigorously it is not (as the integral can be UV divergent). The steps leading to the hypergeometric representation are  
\begin{equation}
\begin{split}
I_{\alpha \beta}^d (m) &=  \frac{\Gamma (\alpha + \beta)}{\Gamma (\alpha) \Gamma (\beta)} \int_0^1 dx x^{\alpha-1} (1-x)^{\beta-1} \int_p \frac{1}{\left[p^2+x m^2+(1-x)M^2 \right]^{\alpha+\beta}}\\
&= \frac{ \Gamma \left(\alpha+\beta-\frac{d}{2} \right)(M^2)^{\frac{d}{2}-\alpha-\beta}}{(4 \pi)^\frac{d}{2} \Gamma (\alpha) \Gamma(\beta)} \int_0^1 dx x^{\alpha-1} (1-x)^{\beta-1} \left[1-\left(1- \frac{m^2}{M^2}\right)x \right]^{\frac{d}{2}-\alpha-\beta}\\
&= \frac{ \Gamma \left(\alpha+\beta-\frac{d}{2} \right)(M^2)^{\frac{d}{2}-\alpha-\beta}}{(4 \pi)^\frac{d}{2} \Gamma (\alpha+\beta)} {}_2 F_1 \left[\alpha+\beta-\frac{d}{2}, \alpha, \alpha+\beta, 1-\frac{m^2}{M^2} \right],
\end{split}
\end{equation}
which can be carried out for $\alpha+\beta-\frac{d}{2}>0$ via standard definitions. By using the well-known hypergeometric series expansion for low variable values, we find the following presentation
\begin{equation}
\begin{split}
I_{\alpha \beta}^d (m,M) &= \frac{ \Gamma \left(\alpha+\beta-\frac{d}{2} \right)(M^2)^{\frac{d}{2}-\alpha-\beta}}{(4 \pi)^\frac{d}{2} \Gamma (\alpha+\beta)} {}_2 F_1 \left[\alpha+\beta-\frac{d}{2}, \alpha, \alpha+\beta, 1-\frac{m^2}{M^2} \right]\\
&=\frac{ \Gamma \left(\alpha+\beta-\frac{d}{2} \right)(M^2)^{\frac{d}{2}-\alpha-\beta}}{(4 \pi)^\frac{d}{2} \Gamma (\alpha+\beta)} \sum_{n=0}^\infty \sum_{k=0}^n \frac{\left(\alpha+\beta-\frac{d}{2} \right)_n  (\alpha)_n}{\left(\alpha+\beta \right)_n n!} {n \choose k} \left(-\frac{m^2}{M^2} \right)^k\\
&= \frac{ (M^2)^{\frac{d}{2}-\alpha-\beta}}{(4 \pi)^\frac{d}{2} \Gamma (\alpha)}\sum_{n=0}^\infty\sum_{k=0}^n \frac{\Gamma \left(\alpha+\beta-\frac{d}{2}+n \right)  \Gamma (\alpha+n)}{\Gamma \left(\alpha+\beta +n\right)} \frac{1}{k! (n-k)!} \left(-\frac{m^2}{M^2} \right)^k.
\end{split}
\end{equation}
However, it is useful to instead consider the integral in broader sense, first limiting to convergent subset (of parameters) and then after suitable expansion, continuing the finite expression over to the complement subset. This structure is more easily recognizable in other regularization schemes, which motivates these steps:
\begin{equation}
\begin{split}
I_{\alpha \beta}^d (m, M) &= \frac{1}{(4 \pi)^\frac{d}{2} \Gamma \left( \frac{d}{2} \right)} \left[ \int_0^{M^2} dt \frac{t^{\frac{d}{2}-1}}{(t+m^2)^\alpha (t+M^2)^\beta} +\int_{M^2}^\infty dt \frac{t^{\frac{d}{2}-1}}{(t+m^2)^\alpha (t+M^2)^\beta} \right]\\
&= \frac{ M^{d-2\beta}}{(4 \pi)^\frac{d}{2}\Gamma \left( \frac{d}{2} \right)} \int_0^1 \frac{dw w^{\frac{d}{2}-1}}{\left( M^2 w + m^2 \right)^\alpha \left( w + 1 \right)^\beta }\\
&-\frac{ M^{d-2\alpha-2 \beta}}{(4 \pi)^\frac{d}{2}\Gamma \left( \frac{d}{2} \right)} \sum_{n,k=0}^\infty {-\alpha \choose n}{-\beta \choose k} \left(\frac{m^2}{M^2}\right)^n  \frac{1}{\frac{d}{2}-\alpha-\beta-n-k}.\\
\end{split}
\end{equation}

\subsection{Gaussian regularization}
The Gaussian suppression is added to the generalized integral of equation (\ref{eq:gen1}) as demonstrated in earlier sections:
\begin{equation}
\begin{split}
\mathfrak{J}_{\alpha \beta}^d (\delta, m, M) &= \frac{1}{(2 \pi)^d} \int d \Omega_d  \int_0^\infty dp p^{d-1}\frac{e^{-\delta p^2}}{(p^2+m^2)^\alpha(p^2+M^2)^\beta}\\
&= \frac{(m^2)^{\frac{d}{2}-\alpha}}{(4 \pi)^\frac{d}{2} \Gamma \left(\frac{d}{2} \right)(M^2)^\beta } \int_0^\infty dw \frac{ e^{- m^2 \delta w} w^{\frac{d}{2}-1}}{\left(1+ \frac{m^2}{M^2}w \right)^\beta(w+1)^\alpha}.
\end{split}
\end{equation}
Here we encounter an obstacle of sorts. There seems not to be a traditional special function representation to the structure as it is (even though it outwardly seems to be a somewhat straightforward generalization). Thus, we start considering it a bit more carefully, and in a manner similar to the one shown above. First let us, re-write the expression using Feynman parametrization for positive parameter values $\alpha, \beta > 0$, 
\begin{equation}
\begin{split}
\mathfrak{J}_{\alpha \beta}^d (\delta, m, M) &=  \frac{\Gamma (\alpha + \beta)}{\Gamma (\alpha) \Gamma (\beta)}  \int_p \int_0^1 dx x^{\alpha-1} (1-x)^{\beta-1} \frac{e^{-\delta p^2}}{\left[p^2+x m^2+(1-x)M^2 \right]^{\alpha+\beta}}\\
&= \frac{ \Gamma \left(\alpha+\beta-\frac{d}{2} \right)}{ \Gamma (\alpha) \Gamma(\beta)} \int_0^1 dx x^{\alpha-1} (1-x)^{\beta-1} \mathfrak{J}_{\alpha + \beta}^d \left[x m^2+(1-x)M^2\right],\\
\end{split}
\end{equation}
where we noted that we can change the order of integration as the integrand is finite in either order (Fubini's theorem). Now, the structure can be decomposed the integral into terms containing regularization scale and the sole term not containing it:
\begin{equation}
\begin{split}
\mathfrak{J}_{\alpha \beta}^d (\delta, m, M) &=\frac{ \Gamma \left(\alpha+\beta-\frac{d}{2} \right)}{ \Gamma (\alpha) \Gamma(\beta)} \int_0^1 dx x^{\alpha-1} (1-x)^{\beta-1} I_{\alpha \beta}^d \left[x m^2+(1-x)M^2\right] + F(\delta, m, M)\\
&= I_{\alpha \beta}^d (m, M) + F(\delta, m, M), 
\end{split}
\end{equation} 
where we have $\left[\mathds{1}-\int d\delta \partial_\delta \right] F(\delta, m, M) = 0$, as $F(\delta, m ,M)$ contains in each term some explicit $\delta$ dependence. Hence, we can write a generalized statement of what we had earlier, such that
\begin{equation}
\begin{split}
\left[\mathds{1}-\int d\delta \partial_\delta \right]\mathfrak{I}_{\alpha\beta}^d (m,M) \longmapsto I_{\alpha \beta}^d (m, M).
\end{split}
\end{equation}
Finally, we consider the remaining parts of parameter space by expanding the integral in earnest. For generic values of $\alpha$ and $\beta$, we can write the following steps  
\begin{equation}
\begin{split}
\mathfrak{J}_{\alpha \beta}^d (\delta, m, M) &= \frac{1}{(2 \pi)^d} \int d \Omega_d  \int_0^\infty dp p^{d-1}\frac{e^{-\delta p^2}}{(p^2+m^2)^\alpha(p^2+M^2)^\beta}\\
&= \frac{M^{d-2\alpha}}{(4 \pi)^\frac{d}{2} \Gamma \left(\frac{d}{2} \right)} \left[ \int_0^1 dw \frac{ e^{- M^2 \delta w} w^{\frac{d}{2}-1}}{\left(m^2+ M^2 w \right)^\beta(w+1)^\alpha}\right.\\
&+ \left. \frac{1}{M^{2 \beta}} \int_1^\infty dt \frac{ e^{- M^2 \delta w} w^{\frac{d}{2}-\alpha-\beta-1}}{\left(\frac{m^2}{M^2 w}+ 1 \right)^\beta \left(1+ \frac{1}{w} \right)^\alpha} \right]\\
&=\frac{M^{d-2\alpha}}{(4 \pi)^\frac{d}{2} \Gamma \left(\frac{d}{2} \right)} \sum_{n=0}^\infty \frac{\left(\delta M^2 \right)^n}{n!} \int_0^1 dw \frac{w^{\frac{d}{2}+n-1}}{\left(m^2+ M^2 w \right)^\beta(w+1)^\alpha}\\
&+\frac{ M^{d-2\alpha-2 \beta}}{(4 \pi)^\frac{d}{2}\Gamma \left( \frac{d}{2} \right)} \sum_{n,k=0}^\infty {-\alpha \choose n}{-\beta \choose k} \left(\frac{m^2}{M^2}\right)^n  \int_1^\infty dw  e^{- M^2 \delta w} w^{\frac{d}{2}-\alpha-\beta-n-k-1}\\
&=\frac{M^{d-2\alpha}}{(4 \pi)^\frac{d}{2} \Gamma \left(\frac{d}{2} \right)} \sum_{n=0}^\infty \frac{\left(\delta M^2 \right)^n}{n!} \int_0^1 dw \frac{w^{\frac{d}{2}+n-1}}{\left(m^2+ M^2 w \right)^\beta(w+1)^\alpha}\\
&+\frac{ M^{d-2\alpha-2 \beta}}{(4 \pi)^\frac{d}{2}\Gamma \left( \frac{d}{2} \right)} \sum_{n,k=0}^\infty {-\alpha \choose n}{-\beta \choose k} \left(\frac{m^2}{M^2}\right)^n  E_{-\frac{d}{2}+\alpha+\beta+n+k+1} \left(\delta M^2 \right),\\
\end{split}
\end{equation}
where the last expression is given in terms of exponential integrals. This result can be related back to the one found in dimensional regularization by considering the small variable expansion of exponential integral:
\begin{equation}
E_\omega(x) = x^\omega \left[\frac{\Gamma (1-\omega)}{x} +\mathcal{O}\left(x^5 \right) \right] +\frac{1}{\omega-1} +\mathcal{O}(x).
\end{equation}
Thus, we can write explicitly
\begin{equation}
\begin{split}
\mathfrak{J}_{\alpha \beta}^d (m, M) &= I_{\alpha \beta}^d (m, M) + G(\delta, m, M),
\end{split}
\end{equation}
where each term in $G(\delta, m, M)$ is dependent on the scale $\delta$. This again verifies the relevant operator identity that takes place when the expression can be Feynman parametrized.
Thus, we can recognize that the series expansions are the finite analytic continuation we were after.
\subsection{UV cut-off}
We again modify the integrand by inserting a suitable Heaviside step function. Thus, our new integral of interest can be re-written such that
\begin{equation}
\begin{split}
\tilde{I}_{\alpha \beta}^d(K, m, M) &= \frac{1}{(2\pi)^d} \int d\Omega_d \int_0^{K} dp \frac{p^{d-1}}{(p^2+m^2)^\alpha(p^2+M^2)^\beta} \\
&= \frac{ K^d}{(4 \pi)^\frac{d}{2}\Gamma \left( \frac{d}{2} \right)} \int_0^1 \frac{dw w^{\frac{d}{2}-1}}{(K^2 w + m^2)^\alpha(K^2 w + M^2)^\beta }\\
&= \frac{ K^d}{(4 \pi)^\frac{d}{2}\Gamma \left( \frac{d}{2} \right)(m^2)^\alpha (M^2)^\beta} \int_0^1 \frac{dw w^{\frac{d}{2}-1}}{\left(\frac{K^2}{m^2} w + 1 \right)^\alpha \left(\frac{K^2}{M^2} w + 1 \right)^\beta }\\
\end{split}
\end{equation}
The remaining integral can be associated with Appell's hypergeometric function for positive parameter values. The explicit identity is given by \cite{bailey}
\begin{equation}
\int_0^1 dx \frac{x^{\alpha-1}(1-x)^{\gamma -\alpha-1}}{(1-xv)^\beta (1-xw)^\delta} = \frac{\Gamma (\alpha) \Gamma (\gamma-\alpha)}{\Gamma(\gamma)} F_1 \left[\alpha, \beta, \delta, \gamma, v, w \right].  
\end{equation}
Thus, we can write the full solution as 
\begin{equation}
\begin{split}
\tilde{I}_{\alpha \beta}^d (K, m,M) = \frac{ K^d}{(4 \pi)^\frac{d}{2}\Gamma \left(1+ \frac{d}{2} \right)(m^2)^\alpha (M^2)^\beta}  F_1 \left[\frac{d}{2}, \alpha, \beta, \frac{d}{2}+1, -\frac{K^2}{m^2}, -\frac{K^2}{M^2} \right].
\end{split}
\end{equation}
While this special function solution is rather elegant looking, the easiest method to extract the dimensionally regularized structure, is again through Feynman parametrization (for convergent $\alpha, \beta > 0$). Thus, performing expansion over $K^2 \gg M^2, m^2$, we are able to change the order of integration, with solely finite integrands in finite ranges. This allows us to write
\begin{equation}
\begin{split}
\tilde{I}_{\alpha \beta}^d (m, M) &= \int_0^1 dx x^{\alpha-1} (1-x)^{\beta-1} \tilde{I}_{\alpha+\beta}^d \left[xm^2+(1-x)M^2 \right]\\
&= I_{\alpha \beta}^d (m, M)+ K^{d-2 \alpha}G(K, m, M),
\end{split}
\end{equation}
where the cut-off has been fully contained in the isolated term on the right-hand side. Let us again note that each term in $ K^{d-2 \alpha}G(K, m, M)$ contains explicit cut-off scale dependence and thus $\left[\mathds{1}- \int dK \partial_K \right] K^{d-2\alpha}G(K,m,M) = 0$. This allows us to perform a same type of generalization as shown in previous section
\begin{equation}
\left[ \mathds{1} - \int dK \partial_K  \right] \tilde{I}_{\alpha \beta}^d (K, m, M) \longmapsto I_{\alpha \beta}^d (m, M).
\end{equation}
However, the results above have been limited to a subset of interesting cases. Hence, we approach the integral through a suitable decomposition and expansion to associate the result explicitly with the sought-after structure via dimensional regularization. The steps are given by
\begin{equation}
\begin{split}
\tilde{I}_{\alpha \beta}^d(K, m, M) &= \frac{1}{(2\pi)^d} \int d\Omega_d \int_0^{K} dp \frac{p^{d-1}}{(p^2+m^2)^\alpha(p^2+M^2)^\beta} \\
&= \frac{ M^{d-2\beta}}{(4 \pi)^\frac{d}{2}\Gamma \left( \frac{d}{2} \right)} \int_0^1 \frac{dw w^{\frac{d}{2}-1}}{\left( M^2 w + m^2 \right)^\alpha \left( w + 1 \right)^\beta }\\
&+\frac{ K^{d-2\alpha-2 \beta}}{(4 \pi)^\frac{d}{2}\Gamma \left( \frac{d}{2} \right)}\int_\frac{M^2}{K^2}^1 dt  \frac{t^{\frac{d}{2}-1}}{\left(t+ \frac{m^2}{K^2} \right)^\alpha \left( \frac{M^2}{K^2}+t \right)^\beta}\\
&= \frac{ M^{d-2\beta}}{(4 \pi)^\frac{d}{2}\Gamma \left( \frac{d}{2} \right)} \int_0^1 \frac{dw w^{\frac{d}{2}-1}}{\left( M^2 w + m^2 \right)^\alpha \left( w + 1 \right)^\beta }\\
&+\frac{ K^{d-2\alpha-2 \beta}}{(4 \pi)^\frac{d}{2}\Gamma \left( \frac{d}{2} \right)} \sum_{n,k=0}^\infty {-\alpha \choose n}{-\beta \choose k} \left(\frac{m^2}{K^2}\right)^n\left(\frac{M^2}{K^2}\right)^k  \frac{1}{\frac{d}{2}-\alpha-\beta-n-k}\left[1-\left(\frac{M^2}{K^2} \right)^{\frac{d}{2}-\alpha-\beta-n-k} \right]\\
&= I_{\alpha \beta}^d (m, M) +\frac{ K^{d-2\alpha-2 \beta}}{(4 \pi)^\frac{d}{2}\Gamma \left( \frac{d}{2} \right)} \sum_{n,k=0}^\infty {-\alpha \choose n}{-\beta \choose k} \left(\frac{m^2}{K^2}\right)^n\left(\frac{M^2}{K^2}\right)^k  \frac{1}{\frac{d}{2}-\alpha-\beta-n-k},
\end{split}
\end{equation}
where we have explicitly isolated the contributions due to the cut-off in a convergent power series. As is obvious, this result further confirms the structure of the mapping. 
\section{External momentum at one-loop}
A natural continuation to the explicit computations is to establish how the generalized integrals treat an external momentum (as an added scale). For dimensional regularization this type of question is trivial, as it is taken care of by merely changing integration variable by a constant vector, $p \mapsto p+ k$, such that
\begin{equation}
\int_p \frac{1}{[(p+k)^2+m^2]^\alpha} \longmapsto I_\alpha^d (m).
\end{equation}
The UV finite schemes can be associated both efficiently and inefficiently. The naive way would involve blindly setting the cut-off with respect to the loop momentum. However, this leads to the need to evaluate $d$-dimensional angular integrals as part of the expansion. This complication can be avoided by applying some thought. We instead use the explicit shape given to us, and use regularization on the propagator such that
\begin{equation}
\begin{split}
\tilde{I}_\alpha^d(K,k,m) = \int_p \frac{\theta \left(K-|p+k| \right)}{\left[(p+k)^2+m^2\right]^\alpha}  \longmapsto \tilde{I}_\alpha^d(K,m),
\end{split}
\end{equation}
where we have performed the change of integration variable as usual.
This is the first indicator on the essence of this regulatory structure being more (efficiently) related to the propagator structure at one-loop level, than the loop momentum. A similar straight-forward interpretation can be found for the Gaussian case, association being tied to propagator and not the loop momentum:
\begin{equation}
\begin{split}
\mathfrak{I}_\alpha^d(\delta,k,m) &= \int_p \frac{e^{-\delta (p+k)^2}}{\left[(p+k)^2+m^2\right]^\alpha}\longmapsto \mathfrak{I}_{\alpha\beta}^d(\delta, m).
\end{split}
\end{equation}
\subsection{External momentum and two mass scales}
While the reasoning given above is convenient for cases with a single propagator structure, upon adding another one, we can no longer find a unique propagator momentum to use in either of the regularization schemes. In this sense consider the first cases with positive $\beta$ and $ \gamma$, which allow us to perform Feynman parametrization on the integrand such that
\begin{equation}
\begin{split}
\int_p \frac{1}{(p^2+m^2)^\beta [(p+k)^2+M^2]^\gamma} &= \frac{\Gamma \left(\gamma+\beta \right)}{\Gamma \left( \beta \right) \Gamma \left(\gamma \right)} \int_{p} \int_0^1 \frac{ dx x^{\gamma-1} (1-x)^{\beta-1}}{\left[(p+xk)^2+ x(1-x)k^2 +x M^2 + (1-x)m^2 \right]^{\gamma+ \beta}}.
\end{split}
\end{equation} 
In a sense we can again see the re-parametrization as a linear combination over proper one-loop propagators, which we in turn regularize such that we add either the exponential suppression or cut-off but now with $p \mapsto p+xk$ to the integrand. Explicitly this reads
\begin{equation}
\begin{split}
\frac{ 1}{\left[(p+xk)^2+ x(1-x)k^2 +x M^2 + (1-x)m^2 \right]^{\gamma+ \beta}} \longmapsto \frac{ \theta \left(K-|p+xk| \right) }{\left[(p+xk)^2+ x(1-x)k^2 +x M^2 + (1-x)m^2 \right]^{\gamma+ \beta}}. 
\end{split}
\end{equation}
This allows us to explicitly change the integration order (once again) and thus achieve the sought-after dimensionally regularized structure to isolate.

Supposing that we are disallowed from using Feynman parameterization, by the merit of having generic values of parameters, our only option is to expand. As the computation is a tiny bit messy, we only perform it for the cut-off regularization. In addition let us ignore the angular part of the integral so far. Effectively, we have now a propagator with angular dependence which we can not fully isolate, and as we perform proper regularization on the angular part, we choose to perform the radial integral first and compare that to what we get with dimensional regularization (i.e. infinity limit with convergent integral). Let us write first
\begin{equation}
\begin{split}
\int_p \frac{1}{(p^2+m^2)^\alpha [(p+k)^2+M^2]^\beta} &\longmapsto J= \int_0^K dp \frac{ p^{d-1}}{(p^2+m^2)^\alpha [(p+zk)^2+k^2(1-z^2)+M^2]^\beta}\\
\end{split}
\end{equation}
where denote the cosine related to the dot product by $z$. The actual computation is achieved by expanding the more convoluted propagator in two different limits using Newton's generalization of binomial theorem. Thus, we find
\begin{equation}
\begin{split}
J&= \sum_{n=0}^\infty \sum_{l=0}^{2n} {-\beta \choose n}{2n \choose l} \frac{(zk)^{2n-l}}{\left[k^2(1-z^2)+M^2\right]^{\beta+n}}\int_0^K dp \frac{ p^{d+l-1} \theta\left(\sqrt{k^2(1-z^2)+M^2}-p-zk \right)}{(p^2+m^2)^\alpha }\\
&+\sum_{n,l=0}^\infty {-\beta \choose n}{-2 \beta-2n \choose l} \frac{\left[k^2(1-z^2)+M^2\right]^{n}}{(zk)^{-l}}\int_0^K dp \frac{ p^{d-2\beta-l-1} \theta\left(p+zk-\sqrt{k^2(1-z^2)+M^2} \right)}{(p^2+m^2)^\alpha }\\
\end{split}
\end{equation}
of which only the latter part contributes to the interesting elements related to the cut-off. The former integral obviously is further cut down by the added Heaviside function. By supposing the latter lower limit non-vanishing and denoting it $A(z) = -zk +\sqrt{k^2(1-z^2)+M^2} > 0$, we can specifically consider
\begin{equation}
\begin{split}
\int_{A(z)}^K dp \frac{p^{d-2\beta-l-1}}{(p^2+m^2)^\alpha} &= m^{d-2\beta-l} \int_{\frac{A(z)}{m}}^{\frac{K}{m}} dt \frac{t^{\frac{d-l}{2}-\beta-1}}{(t+1)^\alpha}\\
&= \theta \left(1-\frac{A(z)}{m} \right) m^{d-2\beta-l} \sum_{j=0}^\infty {-\alpha \choose j} \left[\int_{\frac{A(z)}{m}}^{1} dt t^{\frac{d-l}{2}+j-\beta-1}+\int_1^\frac{K}{m} dtt^{\frac{d-l}{2}-\alpha-j-\beta-1} \right] \\
&+ \theta \left(\frac{A(z)}{m}-1 \right) m^{d-2\beta-l} \sum_{j=0}^\infty {-\alpha \choose j} \int_{\frac{A(z)}{m}}^{\frac{K}{m}} dt t^{\frac{d-l}{2}-\alpha-j-\beta-1}\\
&= -m^{d-2\beta-l} \sum_{j=0}^\infty {-\alpha \choose j} \left\{\frac{\theta \left(1-\frac{A(z)}{m} \right)}{\frac{d-l}{2}+j-\beta} \left[\frac{A(z)}{m} \right]^{\frac{d-l}{2}+j-\beta}\right.\\
&\left.+\frac{\theta \left(\frac{A(z)}{m} -1\right)}{\frac{d-l}{2}-\alpha-j-\beta}\left[\frac{A(z)}{m} \right]^{\frac{d-l}{2}-\alpha-j-\beta}-\frac{\theta \left(1-\frac{A(z)}{m} \right)}{\frac{d-l}{2}+j-\beta} \right\}\\
&+m^{d-2\beta-l} \sum_{j=0}^\infty {-\alpha \choose j} \frac{1}{\frac{d-l}{2}-\alpha-j-\beta}\left(\frac{K}{m} \right)^{\frac{d-l}{2}-\alpha-j-\beta}.
\end{split}
\end{equation}
In this sense, we have established the isolation of the scale even in the case we can not perform regularization efficiently. Hence, we recognize that with the cut-off, the UV region is un-problematic in terms of external momenta. A similar argument can be achieved for Gaussian regularization. At this point we have explored UV behaviour and structures enough to have a sense of how the schemes are implemented and how they relate to the dimensional regularization. What remains is to consider the implementation of scaleless integrals, and treatment of IR divergences.
\section{Single scale IR structure}
In order to consider the vanishing loop integrals via Veltman's identity, we must define an IR finite structure corresponding to the cases we aim to compute. An obvious boon to this is the expansion of parameter space to allow for negative dimensions $d < 0$. As we exclusively discuss integrals of radial type here, these new values of parameters can be interpreted solely through (\ref{eq:axiomgauss}), with said integral acting as an axiom. The sole complication afterwards relates to the convergence of the axiomatic integral, which in turn can be treated by re-defining all radial integrals in the negative (and positive) dimensional axis through either Gaussian or cut-off regularization scheme. This is further discussed in an upcoming article.

  The simplest approach involves recycling the scales we already possess. Specifically we define the finite range integral in terms of the familiar cut-off additionally in the lower end of the integral, such that
\begin{equation}
\begin{split}
\mathfrak{K}_\alpha^d(K,m) &= \frac{2}{(4 \pi)^\frac{d}{2} \Gamma \left( \frac{d}{2} \right)} \int_{\frac{1}{K}}^K dp \frac{p^{d-1}}{\left(p^2+m^2 \right)^\alpha}.\\
\end{split}
\end{equation}
By considering the vanishing mass scale limit $m \rightarrow 0$, the integral becomes
\begin{equation}
\begin{split}
\mathfrak{K}_\alpha^{d}(K) &\equiv \frac{1}{(2 \pi)^d} \int d \Omega_d \int_{\frac{1}{K}}^K dp p^{d-1-2\alpha} \\
&= \frac{2 }{(4 \pi)^\frac{d}{2} \Gamma \left( \frac{d}{2} \right)} \int_0^{\frac{K^2-1}{K}} dv \left(v +\frac{1}{K}\right)^{d-2\alpha-1}\\
&= \frac{2}{d-2\alpha}\frac{1 }{(4 \pi)^\frac{d}{2}} \left[K^{d-2\alpha}-\frac{1}{K^{d-2\alpha}} \right].
\end{split}
\end{equation}
Similarly we can write for the Gaussian integral another IR finite variant. The new prescription could be defined in a somewhat symmetric Gaussian fashion, such that we would replace in the exponent $\delta p^2 \longmapsto \delta \left(p^2 +\frac{1}{p^2} \right)$. However, this leads to somewhat convoluted structures, most of which can not be given in terms of well known special functions. Thus, we instead construct the IR finiteness by performing a suitable lower integration limit (IR cut-off) such that
\begin{equation}
\begin{split}
\mathcal{J}_\alpha^d (\delta, m) &= \frac{2}{(4 \pi)^\frac{d}{2} \Gamma \left( \frac{d}{2}\right)} \int_\delta^\infty dp \frac{e^{-\delta p^2} p^{d-1}}{(p^2+m^2)^\alpha}.\\
\end{split}
\end{equation}
Hence, upon taking the massless limit, we find an expression in terms of confluent hypergeometric function of the second kind:
\begin{equation}
\begin{split}
\mathcal{J}_\alpha^d (\delta) &= \frac{2}{(4 \pi)^\frac{d}{2} \Gamma \left( \frac{d}{2}\right)} \int_\delta^\infty dp e^{-\delta p^2} p^{d-1-2\alpha}\\
&= \frac{e^{-\delta^3}}{(4 \pi)^\frac{d}{2} \Gamma \left( \frac{d}{2}\right)}\int_0^\infty dt e^{-\delta t} \left(t+\delta^2 \right)^{\frac{d}{2}-\alpha-1}\\
&= \frac{\delta^{d-2\alpha} e^{-\delta^3}}{(4 \pi)^\frac{d}{2} \Gamma \left( \frac{d}{2}\right)}\int_0^\infty dv e^{-\delta^3 v} \left(v+1 \right)^{\frac{d}{2}-\alpha-1}\\
&=\frac{\delta^{d-2\alpha} e^{-\frac{\delta^3}{m^2}}}{(4 \pi)^\frac{d}{2} \Gamma \left( \frac{d}{2}\right)}U \left[1, \frac{d}{2}-\alpha, \delta^3\right].\\
\end{split}
\end{equation}
In either case listed above, the expansion of the results is mapped to vanishing value via the familiar scale extraction operator, such that 
\begin{equation}
\left[\mathds{1} - \int dK \partial_K\right] \bar{I}_\alpha^d(K) = \left[\mathds{1} - \int d\delta \partial_\delta\right] \mathcal{J}_\alpha^d(\delta) = 0.
\end{equation}
However, now the results given to the UV finite structures do no longer automatically agree with these limiting finite IR structures. Thus, we need to re-establish the mapping for non-vanishing masses. First note that in case of the cut-off, we can relate the IR finite entity to the familiar UV finite entity such that
\begin{equation}
\begin{split}
\mathfrak{K}_\alpha^d(K,m) &= \frac{2}{(4 \pi)^\frac{d}{2} \Gamma \left( \frac{d}{2} \right)} \int_{\frac{1}{K}}^K dp \frac{p^{d-1}}{\left(p^2+m^2 \right)^\alpha}\\
&= \tilde{I}_\alpha^d(K,m)-\tilde{I}_\alpha^d \left(\frac{1}{K},m \right).
\end{split}
\end{equation}
Here we note that the latter element corresponds to 
\begin{equation}
\begin{split}
\tilde{I}_\alpha^d \left(\frac{1}{K},m \right) &= \frac{ K^{-d}}{(4 \pi)^\frac{d}{2}\Gamma \left(1+ \frac{d}{2} \right)(m^2)^\alpha}   {}_2 F_1 \left[\alpha,\frac{d}{2}, 1+ \frac{d}{2}, -\frac{1}{K^2 m^2} \right]\\
&= \frac{ K^{-d}}{(4 \pi)^\frac{d}{2}\Gamma \left(1+ \frac{d}{2} \right)(m^2)^\alpha}  \left[1 + \mathcal{O} \left( \frac{1}{K^2 m^2} \right) \right].
\end{split}
\end{equation}
Thus, we recognize that $\left[\mathds{1} -\int dK \partial_K \right] \tilde{I}_\alpha^d \left(\frac{1}{K},m \right) = 0$. Hence, we can indeed use this extension to allow for proper IR behaviour, yielding (including Veltman's identity) a generic expression
\begin{equation}
\begin{split}
\left[\mathds{1} -\int dK \partial_K \right] \mathfrak{K}_\alpha^d (K,m) \longmapsto I_\alpha^d(m).
\end{split}
\end{equation}
In case of the mixed Gaussian integral we find instead
\begin{equation}
\begin{split}
\mathcal{J}_\alpha^d(\delta, m) &\equiv  \frac{2}{(4 \pi)^\frac{d}{2} \Gamma \left(\frac{d}{2} \right)} \int_\delta^\infty dp e^{-\delta p^2} \frac{p^{d-1}}{(p^2+m^2)^\alpha}\\
&= \mathfrak{J}_\alpha^d (m) - \frac{1}{(4 \pi)^\frac{d}{2} \Gamma \left(\frac{d}{2} \right)} \int_0^{\delta^2} dt e^{-\delta t} \frac{t^{\frac{d}{2}-1}}{(t+m^2)^\alpha}\\
&=  \mathfrak{J}_\alpha^d (m) - \frac{\delta^{d}m^{-2 \alpha}}{(4 \pi)^\frac{d}{2} \Gamma \left(\frac{d}{2} \right)} \int_0^1 dv  e^{-\delta^3 v}\frac{v^{\frac{d}{2}-1}}{\left( 1+ \frac{\delta^2 }{m^2} v\right)^\alpha}.
\end{split} 
\end{equation}
The remaining integral is not recognizable in terms of any well-known special function. Thankfully we are in position that allows expansion $\left( \frac{\delta^2}{m^2} v \ll 1 \right)$. This, in combination with obvious convergence, allows us to commute the corresponding sum and integral operator. Thus, we recognize
\begin{equation}
\begin{split}
\delta^{d} \int_0^1 dv  e^{-\delta^3 v}\frac{v^{\frac{d}{2}-1}}{\left( 1+ \frac{\delta^2 }{m^2} v\right)^\alpha} &= \sum_{n=0}^\infty {-\alpha \choose n} \frac{\delta^{d+2n}}{m^{2n}} \int_0^1 dv e^{-\delta^3 v} v^{\frac{d}{2}+n-1}\\
&= \sum_{j=0}^\infty \sum_{n=0}^\infty \frac{(-1)^j}{j!} {-\alpha \choose n} \frac{\delta^{d+2n+3j}}{m^{2n}} \int_0^1 dv  v^{\frac{d}{2}+n+j-1}\\
&= \sum_{j=0}^\infty \sum_{n=0}^\infty \frac{(-1)^j}{\left(\frac{d}{2}+n+j \right)j!} {-\alpha \choose n} \frac{\delta^{d+2n+3j}}{m^{2n}}.
\end{split}
\end{equation}
Hence, we can state in good faith that this expression does not contain any terms independent of the scale $\delta$, which in turn means that it vanishes upon being operated on by $\mathds{1}-\int d\delta \partial_\delta$. Thus, the Gaussian expression (as it is given here) can be generalized to take into account the vanishing scaleless integral such that
\begin{equation}
\left[\mathds{1}-\int d\delta \partial_\delta \right] \mathcal{J}_\alpha^d (m) \longmapsto I_\alpha^d(m)
\end{equation}
\subsection{Two-sided gaussian regularization}
For the sake of completeness, let us consider the suggested form of pure Gaussian regularization for either source of divergent behaviour. This reads
\begin{equation}
\begin{split}
\mathcal{G}_\alpha^d (m) &=\int_p \frac{e^{-\delta \left(p^2 +\frac{1}{p^2} \right)}p^{d-1}}{(p^2+m^2)^\alpha}\\
 &= \frac{m^{d-2\alpha}}{(4 \pi)^\frac{d}{2} \Gamma \left( \frac{d}{2} \right)} \left[ \int_0^1 dt \frac{e^{-\delta \left(m^2 t +\frac{1}{m^2 t} \right)}t^{\frac{d}{2}-1}}{(1+t)^\alpha}+ \int_1^\infty dt \frac{e^{-\delta \left(m^2 t +\frac{1}{m^2 t} \right)}t^{\frac{d}{2}-1}}{(1+t)^\alpha} \right]. 
\end{split}
\end{equation}
The former integral can be expanded both in terms of the first exponential function and the denominator such that 
\begin{equation}
\begin{split}
\int_0^1 dt \frac{e^{-\delta \left(m^2 t +\frac{1}{m^2 t} \right)}t^{\frac{d}{2}-1}}{(1+t)^\alpha}  &=\sum_{n,k=0}^\infty {-\alpha \choose n} \frac{\left(-\delta m^2 \right)^k}{k!} \int_0^1 dt e^{-\frac{\delta}{m^2 t}}t^{\frac{d}{2}+n+k-1}\\
&= \sum_{n,k=0}^\infty {-\alpha \choose n} \frac{\left(-\delta m^2 \right)^k}{k!} \int_1^\infty dw \frac{e^{-\frac{\delta w}{m^2}}}{w^{\frac{d}{2}+n+k+1}}\\
&= \sum_{n,k=0}^\infty {-\alpha \choose n} \frac{\left(-\delta m^2 \right)^k}{k!} E_{\frac{d}{2}+n+k+1} \left(\frac{\delta}{m^2} \right),
\end{split}
\end{equation}
where we recognized the last remaining integral as the exponential integral. The latter integral can be considered in similar manner (now over $\frac{1}{t}$ though)
\begin{equation}
\begin{split}
\int_1^\infty dt \frac{e^{-\delta \left(m^2 t +\frac{1}{m^2 t} \right)}t^{\frac{d}{2}-1}}{(1+t)^\alpha}  &=\sum_{n,k=0}^\infty {-\alpha \choose n} \frac{\left(-\frac{\delta}{ m^2} \right)^k}{k!} \int_1^\infty dt e^{-\delta m^2 t}t^{\frac{d}{2}-\alpha-n-k-1}\\
&= \sum_{n,k=0}^\infty {-\alpha \choose n} \frac{\left(-\frac{\delta}{ m^2} \right)^k}{k!} E_{\alpha-\frac{d}{2}+n+k+1} \left(\delta m^2 \right).
\end{split}
\end{equation}
The result is in a sense somewhat unsatisfying. However, we can extract terms without regularization scale dependence by noting that exponential integral can be expanded such that
\begin{equation}
E_\omega(x) = x^\omega \left[\frac{\Gamma (1-\omega)}{x} +\mathcal{O}\left(x^5 \right) \right] +\frac{1}{\omega-1} +\mathcal{O}(x). 	
\end{equation}
Hence for non-integer $\omega$ we recognize that we can isolate the explicit $\delta$ dependence and write
\begin{equation}
\begin{split}
\mathcal{G}_\alpha^d(m) &= \frac{m^{d-2\alpha}}{(4 \pi)^\frac{d}{2} \Gamma \left( \frac{d}{2} \right)} \left[\sum_{n=0}^\infty {-\alpha \choose n} \left(\frac{1}{\frac{d}{2}+n}+\frac{1}{\alpha-\frac{d}{2}+n} \right)+ H(\delta,m)  \right]\\
&= I_\alpha^d(m) + \mathcal{H}(\delta, m).
\end{split}
\end{equation}
Again each term in $\mathcal{H}$ is explicitly dependent on the regularization scale $\delta$, which obviously allows us to write again
\begin{equation}
\begin{split}
\left[\mathds{1}-\int d\delta \partial_\delta \right] \mathcal{G}_\alpha^d (m) \longmapsto I_\alpha^d(m).
\end{split}
\end{equation}
What is more, we can treat with this structure also the IR region in a straightforward manner by recognizing the integral representation of a modified Bessel function \cite{stegun}
\begin{equation}
\begin{split}
\mathcal{G}_\alpha^d(0) &= \int_p e^{-\delta \left(p^2 +\frac{1}{p^2} \right)} p^{d-2 \alpha-1}\\
&= \frac{\delta^{\alpha-\frac{d}{2}}}{(4 \pi)^\frac{d}{2} \Gamma \left( \frac{d}{2} \right)} \int_0^\infty dt e^{ -t-\frac{\delta^2}{t} } t^{\frac{d}{2}-\alpha-1}\\
&=\frac{2}{(4 \pi)^\frac{d}{2} \Gamma \left(\frac{d}{2} \right)}  K_{\alpha-\frac{d}{2}} \left(2 \delta \right).
\end{split}
\end{equation}
For non-integer values of $\alpha-\frac{d}{2}$ we find an expansion, each term of which is dependent on the regularization scale:
\begin{equation}
K_\alpha(x) = x^{-\alpha} \left[2^{\alpha-1} \Gamma (\alpha) +\mathcal{O}(x^2) \right] + x^\alpha \left[ 2^{-\alpha-1} \Gamma \left(-\alpha\right) + \mathcal{O}(x^2)\right].
\end{equation}

 Thus, we can explicitly state that 
\begin{equation}
\left[\mathds{1} -\int d\delta \partial_\delta \right] \mathcal{G}_\alpha^d (0) = 0.
\end{equation}

\section{UV and IR with separate scales}
To find a more general class of descriptors, we need not consider IR and UV scales to be related in any way. While this introduction of additional scale complicates the related algebra a bit, the previous results can be employed to great extent to reduce the workload.
Consider now a case with small $\delta$ and large $K$ as the limits for the integral
\begin{equation}
\begin{split}
\mathcal{P}_\alpha^d (K, \delta, m) &= \frac{2}{(4 \pi)^\frac{d}{2} \Gamma \left( \frac{d}{2} \right)} \int_{\delta}^K dp \frac{p^{d-1}}{(p^2+m^2)^\alpha}\\
&= \tilde{I}_\alpha^d(K,m) -\tilde{I}_\alpha^d(\delta, m) 
\end{split}
\end{equation}
Noteworthy is that the two scales decompose as separate entities, so we can immediately recognize that 
\begin{equation}
\left[\mathds{1} - \int dK \partial_K \right] \mathcal{P}_\alpha^d(K, \delta, m) = I_\alpha^d(m)-\tilde{I}_\alpha^d(\delta, m).
\end{equation}
However, as before, we note that for small $\delta$ we find the expansion to correspond to terms fully dependent on $\delta$. Thus, we can write 
\begin{equation}
\begin{split}
\left[\mathds{1}- \int d\delta \partial_\delta \right] \tilde{I}_\alpha^d (\delta, m) = \left[\mathds{1}- \int d\delta \partial_\delta - \int dK \partial_K \right] \tilde{I}_\alpha^d (\delta, m)=0.
\end{split}
\end{equation}
Similarly we note that 
\begin{equation}
\begin{split}
\left[\mathds{1}- \int d\delta \partial_\delta - \int dK \partial_K \right] \tilde{I}_\alpha^d (K, m)= \left[\mathds{1} - \int dK \partial_K \right] \tilde{I}_\alpha^d (K, m). 
\end{split}
\end{equation}
Thus, we find the operator generalization (corresponding to decomposing scales) as $\mathds{1}- \int dK \partial_K - \int d\delta \partial_\delta$. Suppose that we have terms with both elements (in power series expansion). Now the resulting operator structure will decrease those contributions twice. Thus, we must compensate with additional application of those terms. The ultimate two scale operator structure is thus given by 
\begin{equation}
\mathfrak{O} = \mathds{1} -\int dK \partial_K -\int d\delta \partial_\delta + \int dK \partial_K \int d\delta  \partial_\delta. 
\end{equation}
In straight-forward manner we recognize that by taking the zero mass limit, we find
\begin{equation}
\begin{split}
\mathcal{P}_\alpha^d(K, \delta) &= \frac{2 }{(4 \pi)^\frac{d}{2} \Gamma \left( \frac{d}{2} \right)} \int_0^{K-\delta} dv \left(v +\delta\right)^{d-2\alpha-1}\\
&= \frac{2}{d-2\alpha}\frac{1 }{(4 \pi)^\frac{d}{2}} \left[K^{d-2\alpha}-\delta^{d-2\alpha} \right].
\end{split}
\end{equation}
which leads to
\begin{equation}
\mathfrak{O} \mathcal{P}_\alpha^d(K, \delta) = 0,
\end{equation}
as it should.

 Returning to a more complex structure, let us consider the two-sided Gaussian regulator. This time we add two separate scales to it such that
\begin{equation}
\begin{split}
\mathcal{W}_\alpha^d (\delta, \xi, m) &= \int_p \frac{e^{-\delta p^2 - \frac{\xi}{p^2} }p^{d-1}}{(p^2+m^2)^\alpha}\\
 &= \frac{m^{d-2\alpha}}{(4 \pi)^\frac{d}{2} \Gamma \left( \frac{d}{2} \right)} \left[ \int_0^1 dt \frac{e^{-\delta m^2 t -\frac{\xi}{m^2 t} }t^{\frac{d}{2}-1}}{(1+t)^\alpha}+ \int_1^\infty dt \frac{e^{-\delta m^2 t -\frac{\xi}{m^2 t} }t^{\frac{d}{2}-1}}{(1+t)^\alpha} \right] \\
 &= \frac{m^{d-2\alpha}}{(4 \pi)^\frac{d}{2} \Gamma \left( \frac{d}{2} \right)} \left[\sum_{n,k=0}^\infty {-\alpha \choose n} \frac{\left(-\delta m^2 \right)^k}{k!} E_{\frac{d}{2}+n+k+1} \left(\frac{\xi}{m^2} \right) \right.\\
 &+ \left. \sum_{n,k=0}^\infty {-\alpha \choose n} \frac{\left(-\frac{\xi}{ m^2} \right)^k}{k!} E_{\alpha-\frac{d}{2}+n+k+1} \left(\delta m^2 \right)\right].\\
 &= I_\alpha^d(m) + \mathcal{M}(\delta, \xi, m),
\end{split}
\end{equation}
where we have extracted the sought-after dimensionally regulated result, similar to previous section, from the whole. The remainder obviously combines the two scales at will, and thus justifies the use of ultimate two scale operator structure, in order to extract
\begin{equation}
\mathfrak{O} \mathcal{W}_\alpha^d(\delta, \xi, m) = I_\alpha^d(m).
\end{equation}
Just for completeness, we write explicitly the zero mass limit of the this expression, such that
\begin{equation}
\mathcal{W}_\alpha^d (\delta, \xi) = \frac{\delta^{\alpha-\frac{d}{2}}}{(4 \pi)^\frac{d}{2} \Gamma \left( \frac{d}{2} \right)} K_{\alpha-\frac{d}{2}} \left(2 \sqrt{\xi \delta} \right).
\end{equation}
Obviously, using similar arguments to those in earlier section, we can write $\mathfrak{O} \mathcal{W}_\alpha^d (0) = 0$.

\section{Conclusions}
While highly successful (in book keeping), dimensional regularization tends to be described in somewhat heavy handed manner. The obscurity (and its success) both rely in the implicit (and unrigorous) assumption of convergence. In this report we set out to express the analytically continued structure as a truly finite entity in all of parameter space. While the results shown here relate strongly to other interpretations involving discarded divergent monomials, we established clear description on the injective mapping, in terms of two separate steps: 
\begin{equation}
\int_p \frac{1}{(p^2 + m^2)^\alpha} \overset{\text{UV}}{\longmapsto} \left\{\tilde{I}_\alpha^d(K,m), \mathfrak{J}_\alpha^d(\delta, m) \right\} \overset{\text{UV}+\text{IR}}{\longmapsto} \left\{\mathcal{P}_\alpha^d(K,\delta,m), \mathcal{W}_\alpha^d (\delta, \xi, m)\right\} \overset{\mathfrak{O}}{\longmapsto} I_\alpha^d (m).
\end{equation}
The two schemes described, cut-off and Gaussian, are not unique, but rather two representatives of a class, specifically described by the separation operator $\mathfrak{O}$ and the one-loop master formula. 
This operator structure can be trivially extended to an arbitrary number of scales, by following trivial combinatorics, denoting the advanced removal operator as 
\begin{equation}
\begin{split}
\mathfrak{A} &= \mathds{1} + \sum_{k=1}^{n} \frac{(-1)^k}{k!} \sum_{ \{a_i \neq a_j \}} \int dK_{a_1} \partial_{K_{a_1}}...\int dK_{a_k} \partial_{K_{a_k}}\\
&\equiv \prod_{k=1}^n \left[\mathds{1}-\int dK_{a_k} \partial_{K_{a_k}} \right].
\end{split}
\end{equation}
While this structure holds no value at single one-loop integrals, it serves as an indicator how we would delve towards multi-loop computations. Each loop order would add up to two new scales, with separable integrals as an obvious example of sub-class that functions under the operator structure as desired. 

Also, the review of external impulses and additional mass scales in one-loop integrals opens path to the explicit computation of multi-loop integrals outside of their region of convergence. As a specific example we present a two-loop integral with three positive parameters $\{\alpha, \beta, \gamma \}$. Using the cut-off strategy presented in previous sections we can immediately write it in terms of Feynman parametrization and proper UV and IR regulators:

\begin{equation}
\begin{split}
\int_p \frac{1}{(p^2+m_1)^\alpha} \int_q \frac{1}{(q^2+m_2^2)^\beta [(p+q)^2+m_3^2]^\gamma} &\longmapsto \frac{\Gamma \left(\gamma+\beta \right)}{\Gamma \left( \beta \right) \Gamma \left(\gamma \right)} \int_0^1  dx x^{\gamma-1} (1-x)^{\beta-1} \int_p  \frac{\theta \left(K_1-p \right) \theta \left(p-\frac{1}{K_1}\right)}{(p^2+m_1)^\alpha} \\
&\times \int_{q} \frac{ \theta \left(K_2-|q+xp| \right) \theta \left(|q+xp|-\frac{1}{K_2} \right) }{\left[(q+xp)^2+ x(1-x)p^2 +x m_3^2 + (1-x)m_2^2 \right]^{\gamma+ \beta}}.
\end{split}
\end{equation}

\section*{Acknowledgements}
The author wishes to thank York Schröder, Saga Säppi and Aleksi Vuorinen for enlightening discussions and feedback. The author acknowledges financial support from the Vilho, Yrjö and Kalle Väisälä Foundation of the Finnish Academy of Science and Letters. This work has been additionally supported by the European Research Council, grant no. 725369, and by
the Academy of Finland, grant no. 1322507.

\bibliographystyle{unsrt}
\bibliography{referencespade}
\end{document}